\documentclass[prl,twocolumn,showpacs]{revtex4-1}
\usepackage{amsmath,amssymb}
\usepackage{graphicx,epsfig,color}
\usepackage{times}

\newcommand{\bra}[1]
{ \langle #1 |}
\newcommand{\ket}[1]
{| #1 \rangle}
\newcommand{\aver}[1]
{\left \langle #1 \right \rangle}

\newcommand{\Imm}[1]
{\textrm{Im}\left [ #1 \right ]}
\newcommand{\Ree}[1]
{\textrm{Re} \left [ #1 \right ]}
\newcommand{\Heff}{\mathcal{H}_{\mathrm{eff}}}

\newcommand{\Pgoe}{\mathcal{P}^{GOE}_M(X)}

\begin{document}

\title{Statistics of resonance states in a weakly open chaotic cavity}

\author{Charles Poli}
\author{Olivier Legrand}
\author{Fabrice Mortessagne}
\affiliation{Laboratoire de Physique de la Mati\`ere Condens\'ee, CNRS UMR 6622
\\Universit\'e de Nice-Sophia Antipolis - 06108 Nice cedex 2, France}

\date{August 6, 2010}

\begin{abstract}
In this letter, we demonstrate that a non-Hermitian Random Matrix description can account for both spectral and spatial statistics of resonance states in a weakly open chaotic wave system with continuously distributed losses. More specifically, the statistics of resonance states in an open 2D chaotic microwave cavity are investigated by solving the Maxwell equations with lossy boundaries subject to Ohmic dissipation. We successfully compare the statistics of its complex-valued resonance states and associated widths with analytical predictions based on a non-Hermitian effective Hamiltonian model defined by a finite number of fictitious open channels. 
\end{abstract}
\maketitle

Examples of waves in enclosures can be found in as diverse contexts as room acoustics, guided optics, vibrations of structures, etc. While these situations imply different physical mechanisms, they all belong to the domain of Wave Chaos, which reveals universal features of cavities with non trivial geometries \cite{Sto99}.
As long as these systems can be considered as closed, their generic spectral and spatial properties are currently well described through the theory of Hermitian random matrices from a statistical point of view. Nevertheless, no realistic system is truly closed, thus calling for a description of the coupling mechanisms to the environment. Physically, the latter include bulk absorption, leads or waveguides, as well as dissipative or radiative boundaries. 
In the domain of Wave Chaos, open systems are actively investigated both from experimental and theoretical points of view (see Refs \cite{Kuh05,Fyo05} for reviews). Among the domains concerned by experimental studies, one can cite: microwave cavities \cite{Kuh07a}, optical microcavities \cite{Sch09}, and elastodynamics \cite{Kuh05,Xer09}.

To analyze these open chaotic systems, the scattering approach was found to be a powerful theory \cite{Fyo05}. In this framework, a system composed of $N$ resonances is analyzed in terms of a $N\times N$ non-Hermitian random matrix, the so-called effective Hamiltonian: 
\begin{equation}
 \Heff = H - \frac{i}{2}VV^{\dagger}\, ,
\end{equation}
  where its Hermitian part $H$ corresponds to the Hamiltonian of the closed system and its anti-Hermitian part $-\frac{i}{2}VV^\dag$ models the coupling to the environment. More precisely, the openness is introduced by means of the $N\times M$ coupling matrix $V$ whose elements $V_n^j$ connect the $n=1,\dots ,N$ states to the $j=1,\dots, M$ scattering channels \cite{Sto99}. The non-Hermiticity of $\Heff$ yields a set of complex eigenvalues $\{ \mathcal{E}_n  \}$ associated to two distinct sets of eigenvectors called left $\{ \bra{\tilde{\psi}_n} \}$ and right $\{  \ket{\psi_n}  \}$ eigenvectors: 
\begin{equation}
\Heff \ket{ \psi_n} = \mathcal{E}_n\ket{ \psi_n}\,, \quad   \bra{\tilde{\psi}_n}\Heff  = \bra{\tilde{\psi}_n} \mathcal{E}_n\, ,
\end{equation}
 where the eigenvalue $\mathcal{E}_n=E_n-i\Gamma_n/2$ gives respectively the energy  $E_n$ and the resonance width $\Gamma_n$ of the $n$th resonance. The left and right eigenvectors, which describe the resonance states respectively for systems  with positive or negative gain, form a bi-orthogonal and complete set which can be normalized by: $\langle \tilde{\psi}_n | \psi_n \rangle = \delta_{nm}$. 

\begin{figure}[t]
\begin{center}
\includegraphics[width=1.65in]{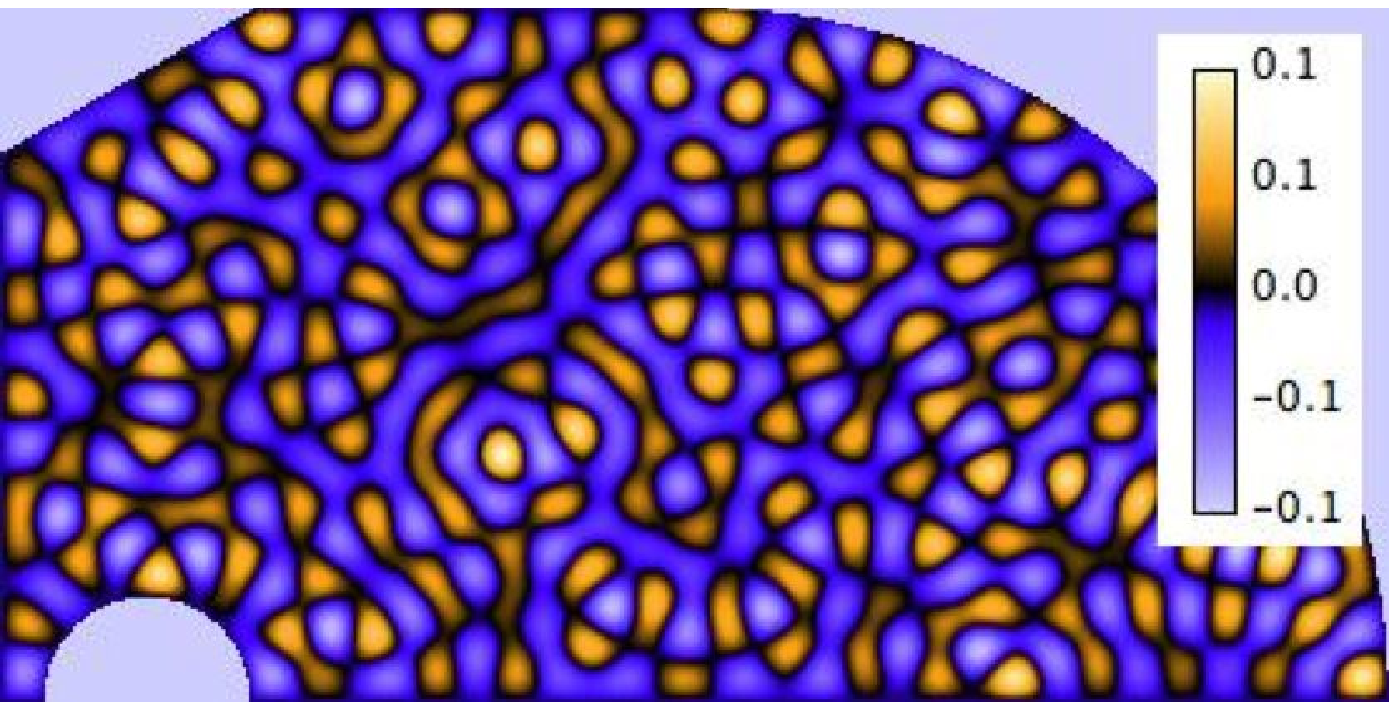}
\includegraphics[width=1.65in]{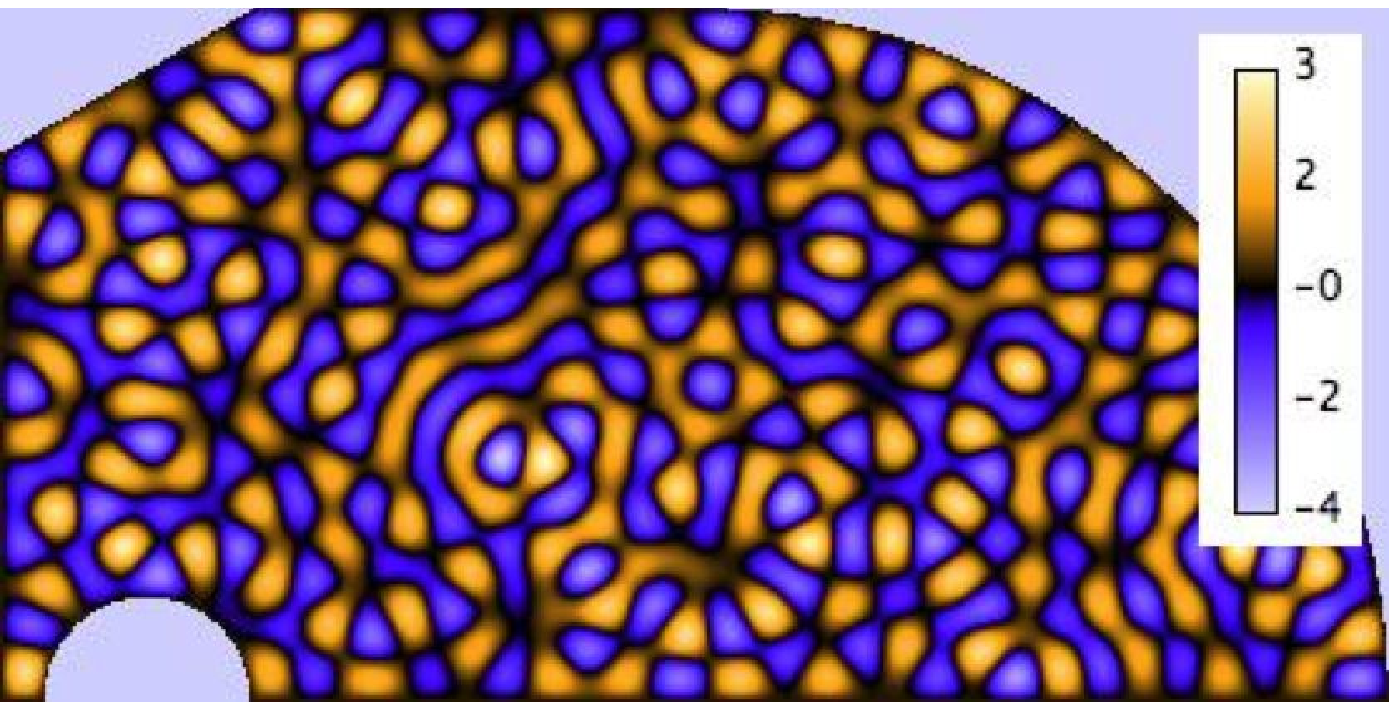}
\end{center}
\caption{Real (on the left) and imaginary (on the right) component of the 500th resonance state obtained by means of the Finite Element Method.}
\label{CavBF}
\end{figure}   

While field statistics of open chaotic systems have been systematically studied for a given energy and considering the energy as a continuous parameter (see \cite{Kuh07a} and references therein), statistics of resonance states \textit{i.e.} left and right eigenvectors of $\Heff$, for TRS systems are less understood. For these systems, the impact of the openness is to turn real eigenfunctions into complex internal wavefunctions associated to resonances. This complexness being uniquely related to the presence of currents inside the system \cite{Pni96,Kim05} (after a phase rotation leading to independent real and imaginary components  \cite{Sai02}). In order to quantify the presence of currents, one can use the complexness parameter $q^2_n$ introduced by Lobkis and Weaver \cite{Lob00} as the ratio of the variance of the imaginary and real parts of the $n$th resonance state. 

Making use of right eigenvectors: 
\begin{equation}\label{q2def}
q_n^2 = \frac{\sum_i(\Imm{\psi_n^i})^2}{\sum_i (\Ree{\psi_n^i})^2}\, ,
\end{equation}
where $\psi_n^i$ corresponds to the $i$th component of the right eigenvector (we note that $q_n^2$ can also be defined using left eigenvectors). This parameter has recently  gained attention both experimentally \cite{Bar05b,Xer09} and theoretically \cite{Sav06,Pol09b} in particular due to its relationship to the resonance width. A linear relationship between $q_n$ and $\Gamma_n$ was first noticed by Barth\'elemy \textit{et al.}  \cite{Bar05b} analyzing hundreds of resonance states of a 2D chaotic microwave cavity at room temperature. This result was then confirmed using the effective Hamiltonian formalism in the limit $M\gg1$, relevant in the experiment \cite{Sav06}. A linear relationship between $q_n$ and $\Gamma_n$ was also verified in an elastodynamics experiment for a given resonance when a spatially extended coupling is varied \cite{Xer09}. Lately the complexness parameter was investigated at arbitrary $M$ by means of its probability distribution in the regime of weak coupling  \cite{Pol09b}. There, it was shown that the average value of $q_n^2$ is directly proportional to the variance of $\Gamma_n$, which constitutes the natural measure of the fluctuations of the widths. 

It is the aim of the present work to confront our theoretical predictions to numerical solutions of the Maxwell equations in a 2D chaotic microwave cavity with lossy boundaries subject to Ohmic dissipation. After a brief introduction to the theoretical model \cite{Pol09b}, we will describe the cavity we numerically investigate.Then we will discuss the statistical results we obtained concerning the complex wavefunctions and the widths of its resonance states and will compare them to the theory.

In the regime of weak coupling, which was shown \cite{Pol09T} to correspond to the condition $\sqrt{\text{var} (\Gamma) } \ll \Delta$, where $\Delta$ is the mean level spacing, the anti-Hermitian part of $\Heff$ is small compared to the Hermitian part and the perturbation theory can be applied. As the eigenvalues are nondegenerate with the probability 1 due to the linear level repulsion at small spacings \cite{Sto99}, one gets directly expressions of the spectral widths and the complexness parameters (\ref{q2def}):
\begin{equation}\label{Gq2}
\Gamma_n =\sum_{j=1}^M(V_n^j)^2\, , \quad q_n^2= \sum_{p\neq n}\frac{\Gamma_{np}^2}{(E_n-E_p)^2} \,,
\end{equation}
where $\Gamma_{np}=\sum_{j=1}^MV_n^jV_p^j$ and the $\{E_n \}$ correspond to the energies of the closed system. We would like to stress that $q_n^2$ also gives information about the non-orthogonality of eigenfunctions, with important implications in various physical situations. For example, the non-orthogonality induces an enhancement of the line width of a lasing mode \cite{Sch00}, influences branching ratios of nuclear cross sections \cite{Sok97} and is also investigated in open quantum maps \cite{Sch04,Kea08}. The distribution of the energies $\{E_n\}$ corresponds to the eigenvalue distribution of the Gaussian Orthogonal Ensemble with $N\rightarrow \infty$ \cite{Sto99}. The coupling amplitudes are chosen to be real Gaussian random variables with zero mean and covariance $\aver{V_n^jV_m^k}= \sigma^2 \delta_{nm}\delta^{jk}$ \cite{Sok89}. To obtain an analytical expression of the distribution, we use the fact that $\Gamma_{np}=\sum_{j=1}^MV_n^jV_p^j$ can be viewed as a scalar product between $M$-dimensional vectors and then can be expressed using polar variables  \cite{Sok89}: $\Gamma_{np}=\sqrt{\Gamma_n\Gamma_p}\cos \theta_{np}$. The complexness parameter is now given by:  
\begin{equation}
q^2_n=\Gamma_n\sum_{p\ne n}\frac{A_{np}}{4(E_n-E_p)^2}\, ,
\end{equation}
where $A_{np}=\Gamma_p\cos^2 \theta_{np}$. Considering that the widths are given by a sum of $M$ squared independent Gaussian random variables (\ref{Gq2}), the distribution of the rescaled widths $\gamma=\Gamma/\sigma^2$ (with $\aver{\gamma}=M$) is given by a $\chi^2$ distribution with $M$ degrees of freedom:

\begin{equation}\label{pg}
  \chi^2_M(\gamma) = \frac{1}{2^{M/2}\Gamma(M/2)}\gamma^{M/2-1}e^{-\gamma/2}\,,
\end{equation}
and a further calculation shows that $A$ satisfies the Porter-Thomas distribution \cite{Pol09b}: $P(A/\sigma^2)=\chi^2_{M=1}$ which  is independent of $M$. The distribution of the rescaled parameter $X_n\equiv \frac{\sigma^4}{\Delta^2}q^2_n$, is defined by $\Pgoe=\aver{\delta(X-X_n)}$, where the statistical averages are performed over the energies $\{E_n\}$, the resonances widths $\{\Gamma_n\}$ and the $\{A_n\}$. By making use of group integral methods \cite{Sch00} it reads:
\begin{equation}
\label{pMX}
 \Pgoe=\frac{\pi^2M}{24X^2}\frac{1+\frac{\pi^2(3+M)}{4X}}{(1+\frac{\pi^2}{4X})^{M/2+2}} \, ,
\end{equation}
where strong mode-to-mode fluctuations clearly appear, embodied in the power law tail $1/X^2$ of the distribution.

In the following, this prediction will be compared with the numerical solutions of the Maxwell equations in a 2D chaotic microwave cavity.  The chaotic cavity we consider has a quarter of a stadium shape with a radius of $R=1$\,m and a length of $l=2$\,m (see Fig. \ref{CavBF}). In order to reduce the bouncing ball modes between the two parallel sides, an oblique cut is performed on one side and a movable perfectly reflecting half disk of diameter $d=0.3$\,m is placed on the opposite side. The absorbing boundary condition is imposed on an adjustable part of the upper arc of circle of the cavity. For the TM polarization, the electromagnetic field $(\textbf{E},\textbf{H})$ inside the cavity is uniquely characterized by the single component of the electric field: $\psi(\textbf{r})=E_z(\textbf{r})$, where $\textbf{r}=(x,y)$. The field component is solution of the  Helmholtz equation: 
\begin{equation}\label{Helm}
-\Big(\frac{\partial^2}{\partial x^2}+\frac{\partial^2}{\partial y^2}\Big)\psi(\textbf{r})=\frac{\tilde{\omega}^2}{c^2}\psi(\textbf{r}) \, ,
\end{equation}
 where $\tilde{\omega}$ is a complex angular frequency and $c$ the light velocity. Starting from the ideal closed system where the field satisfies the Dirichlet boundary condition \textit{i.e.} an infinite conductivity along the whole contour of the cavity: $\psi^{(0)}|_\mathcal{C}=0$, the openness is introduced through a finite conductivity $\sigma_c$ on a length $l_{abs}$ of the contour. To first order, the electric field $\psi$ along the lossy contour is then given by: 
 \begin{equation}
\label{mixedbc}
\psi \simeq -(1+i) \sqrt{\dfrac{1}{2\mu_0\sigma_{c}\omega}}\,\hat{n}\cdot \vec{\nabla}\psi^{(0)}\,,
\end{equation}
where $\sigma_{c}$ is the effective conductivity of the contour, and $\hat{n}$ is the unit normal vector directed toward the interior of the conductor \cite{Jac98, Bar05b}. 
 
As we examine resonance states, the Finite Element Method  \cite{Jin93} reveals to be very efficient to solve the time independent wave equation (\ref{Helm}). Using the commercial software Comsol$^\text{TM}$, we are able to obtain the complex eigenvalues $\tilde{\omega}_n=\omega_n-i\zeta_n/2$ and the resonance states $\psi_n$ of the first 800 resonances. To increase the statistical samples at our disposal, ensemble averages were performed by sliding the half disk reflector along the largest side of the cavity (ensuring a constant area and perimeter of the cavity) with 7 different positions well enough separated to produce completely statistically independent spectra leading to samples of 2100 resonances for each numerical distribution. To compare the numerical results from the electromagnetic cavity with the theoretical model, the usual correspondence between ($E_n, \Gamma_n$) and ($\omega_n^2,\omega_n \zeta_n$) is performed.

In the 2D chaotic cavity we study, the number of channels $M$ is related to
the effective absorbing length $l_{abs}$ along the boundary. This number can
be evaluated by using the Sabine's law of reverberation known in room
acoustics \cite{Sav06}. This law is formally equivalent to the
so-called Weisskopf's estimate for the level width (well-known in Nuclear
Physics \cite{Sok89}) provided the effective number of channels be
related to the absorbing perimeter through the intuitive relationship :
\begin{equation}\label{Sab}
M=\frac{l_{abs}}{\lambda/2} ,
\end{equation}
where $\lambda = 2 \pi c / \omega$ is the wavelength. Clearly, as this
estimate is frequency dependent, to compare numerical distributions to our
theoretical predictions, one must use samples of resonances for which the
parameter $M$ is approximately a constant. This can be achieved by
considering resonances within high-frequency intervals. Indeed, as the
cumulated number of levels grows like $\omega^2$ and $M$ like $\omega$ (according to (\ref{Sab})), the relative variation of $M$ within a sequence of $\Delta N$ adjacent levels
around the $N$th level is given by $\Delta M / M = \frac{1}{2} \Delta N /
N$. In practice, we considered intervals of 100 adjacent resonances above the
300th resonance. To explore various numbers of channels, namely two
frequency ranges were considered : from the 300th to the 400th resonances
($\Delta M / M \simeq 14$\%), and from the 700th to the 800th resonance
($\Delta M / M \simeq 7$\%), and three different lengths of the absorbing part of the perimeter : $l_{abs}=\pi/18$, $l_{abs}=\pi/6 $ and $l_{abs}=\pi/2$ (in unit of radius). It is also important to note that the theoretical distribution of the complexness parameter (\ref{pMX}) being established in the regime of weak coupling, the conductivity of the absorbing part of the perimeter was chosen such that $\sqrt{\textrm{var}(\Gamma)}/\Delta < 10^{-2}$.

\begin{figure}[t]		
\begin{center}
\includegraphics[width=2.5in]{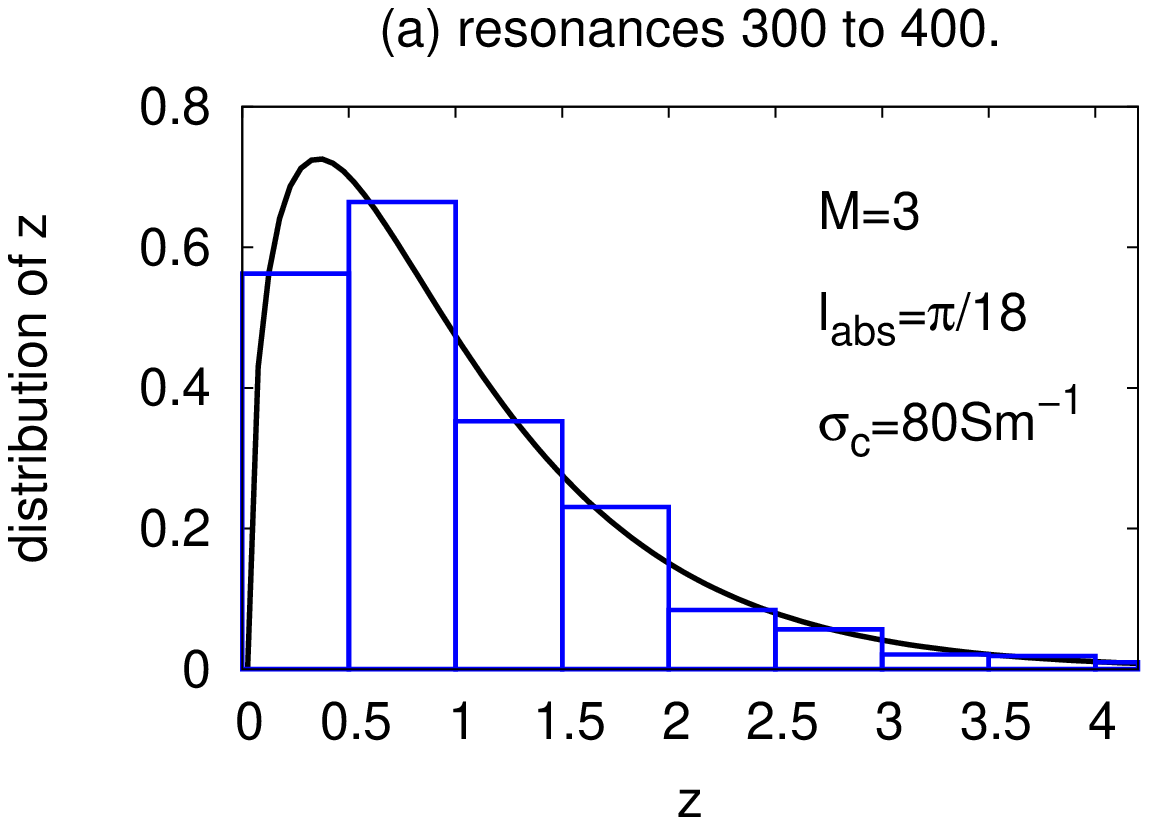}\\
\includegraphics[width=2.5in]{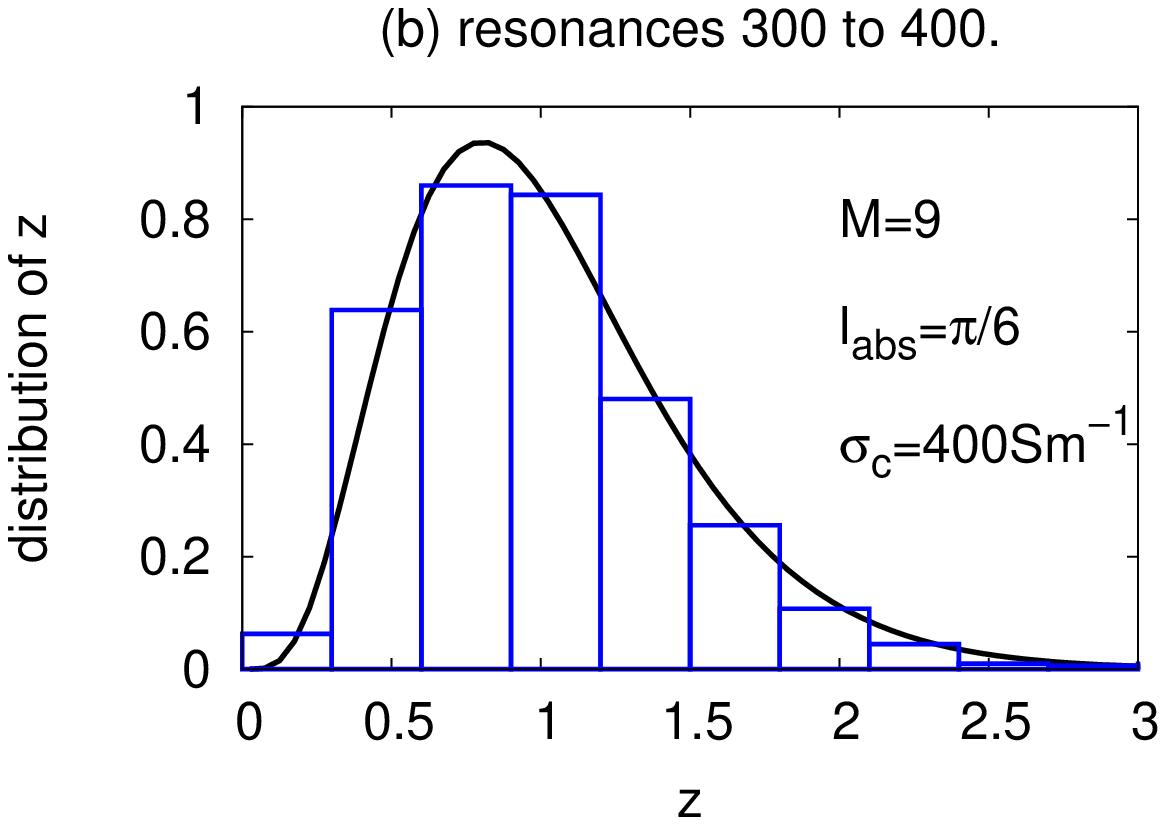}\\
\includegraphics[width=2.5in]{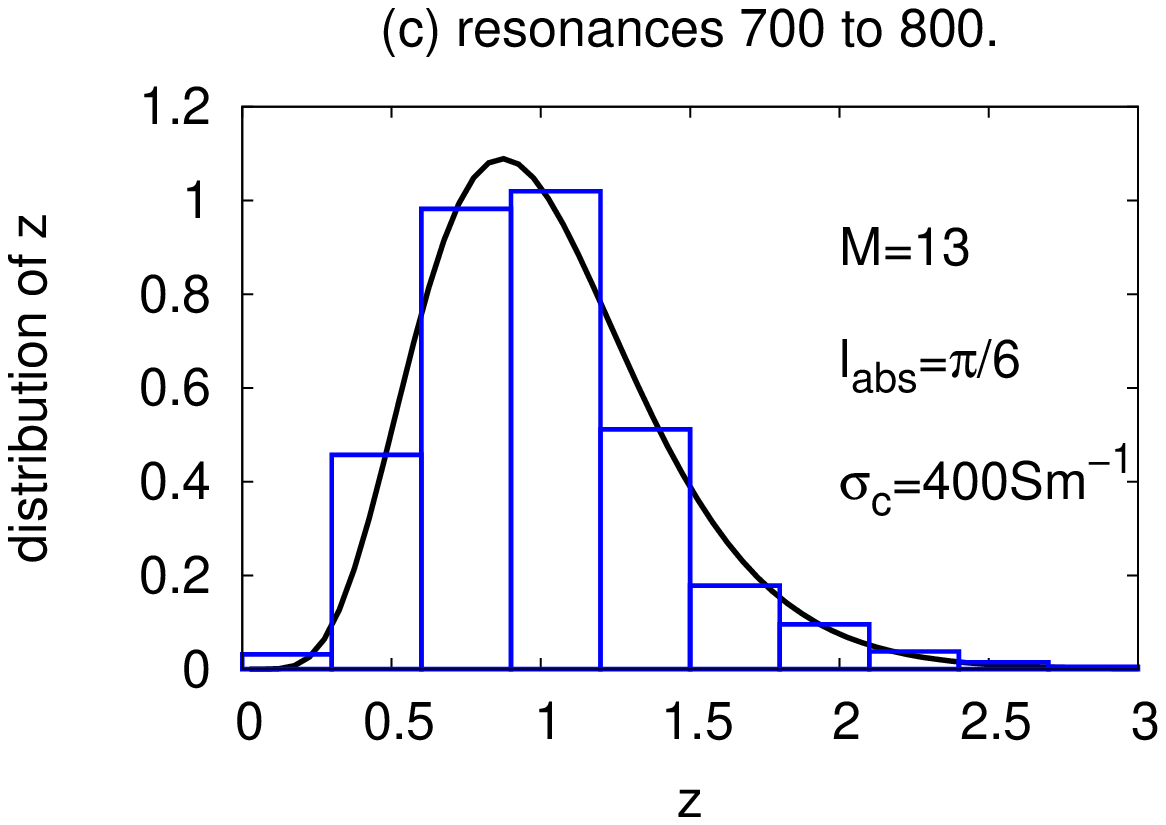}\\
\includegraphics[width=2.5in]{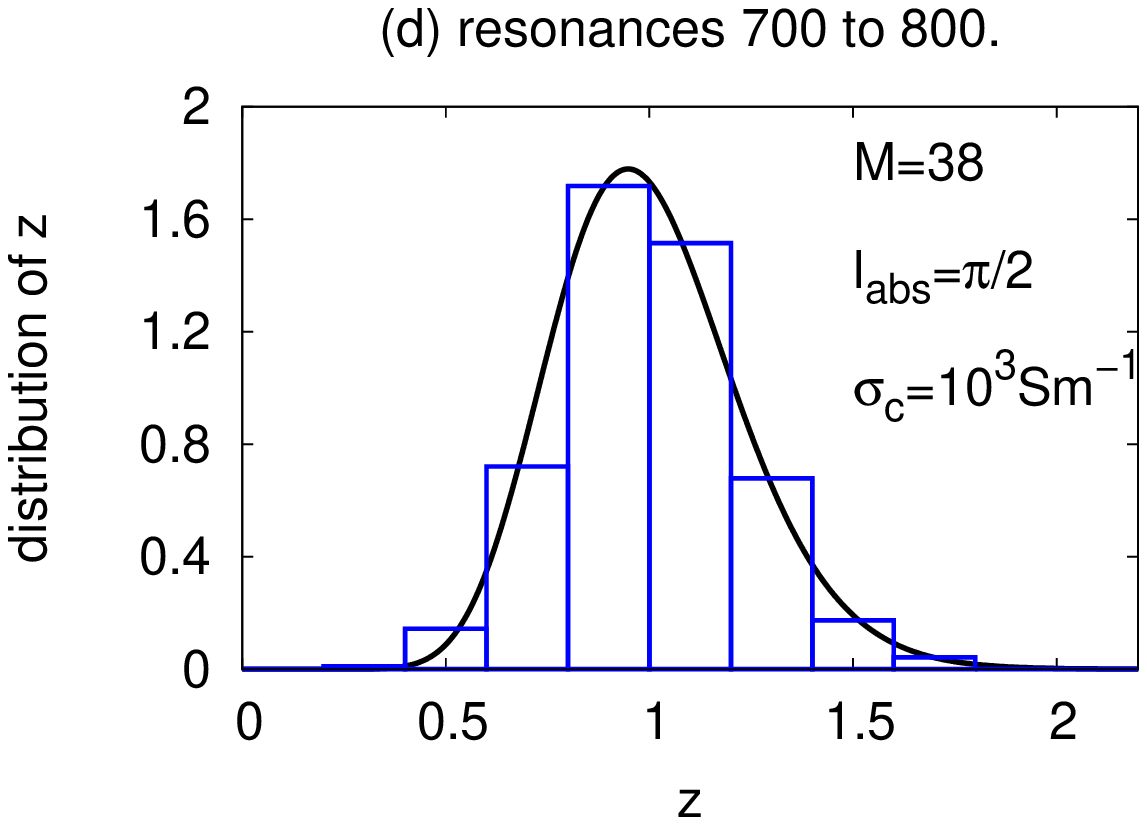}
\end{center}
\caption{(Color on line). Distributions of the widths $z=\gamma/\aver{\gamma}$ for different absorbing lengths, conductivities and frequency ranges: (a)   $l_{abs}=\pi/18$, $\sigma_c=80$Sm$^{-1}$ and from the 300th to the 400th resonances;  (b)   $l_{abs}=\pi/6$, $\sigma_c=400$Sm$^{-1}$ and from the 300th to the 400th resonances; (c)   $l_{abs}=\pi/6$, $\sigma_c=400$Sm$^{-1}$ and from the 700th to the 800th resonances; (d)   $l_{abs}=\pi/2$, $\sigma_c=1000$Sm$^{-1}$ and from the 700th to the 800th resonances. The histograms show the numerical distribution. The solid lines correspond to the $\chi^2_M$  law (\ref{pg}). The number $M$ of channels used for the comparison corresponds to the nearest integer value of relation (\ref{Sab}) computed with the median value of the wavelength in each frequency interval.}
\label{pgFig} 
\end{figure}

Fig. \ref{pgFig} shows the distributions of the spectral widths  $z=\gamma/\aver{\gamma}$ of the cavity compared to the $\chi^2$ distribution (\ref{pg}). 
Four different sets of values of the absorbing lengths, conductivities and frequency ranges are used (see caption). For each interval of frequency, the number $M$ of channels used for the comparison corresponds to the nearest integer value of relation (\ref{Sab}) computed with the median value of the wavelength in each interval. The remarkable agreement demonstrates that the concept of effective independent coupling channels is physically relevant to describe continuously distributed losses. 
Note that the number of channels can also be obtained through the first two moments of the width distribution (\ref{pg}): 
\begin{equation}\label{Mom}
M=\frac{2}{\aver{\gamma^2}/\aver{\gamma}^2-1} \, .
\end{equation}
We checked that except for the smallest value $M=3$ the estimates (\ref{Sab}) and  (\ref{Mom}) agree within a few percents. In the worst case ($M=3$) the discrepancy is only of 17\%, consistent with the relative variation of $M$ given above. 

\begin{figure}[t]
\begin{center}
\includegraphics[width=2.5in]{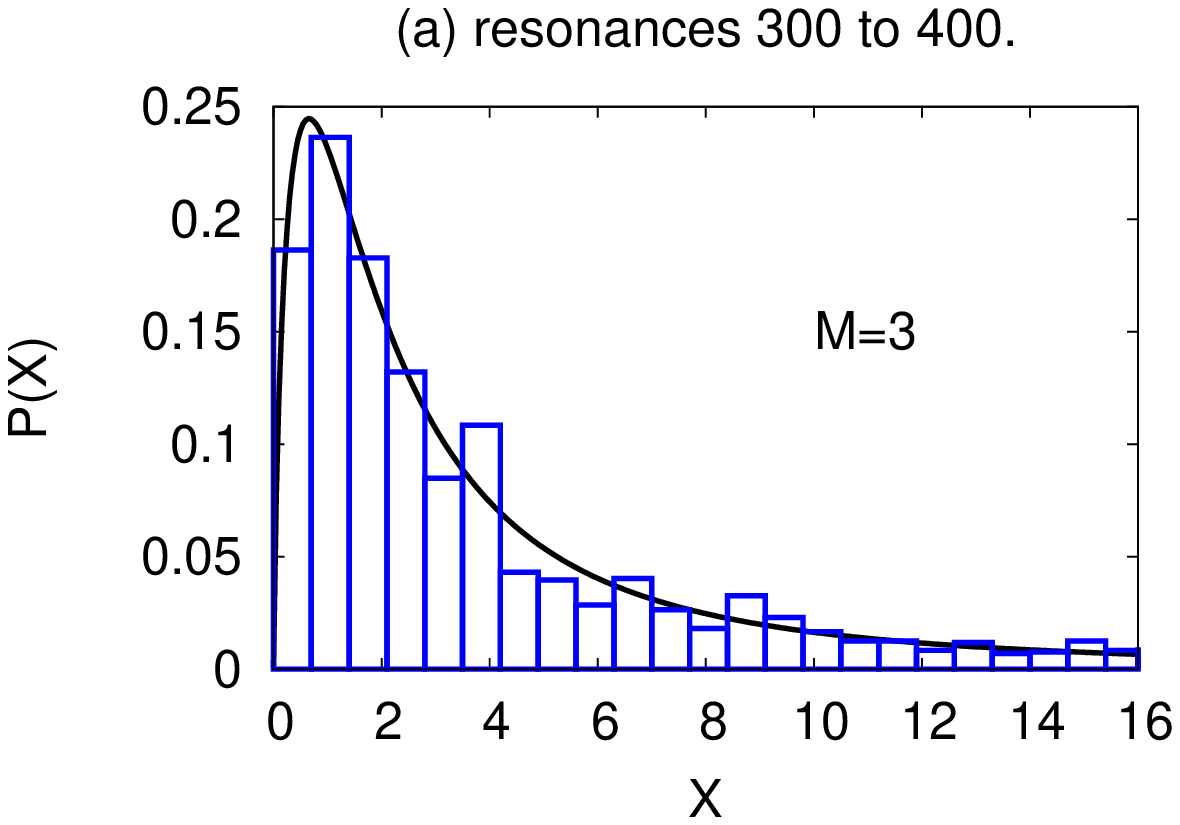}\\
\includegraphics[width=2.5in]{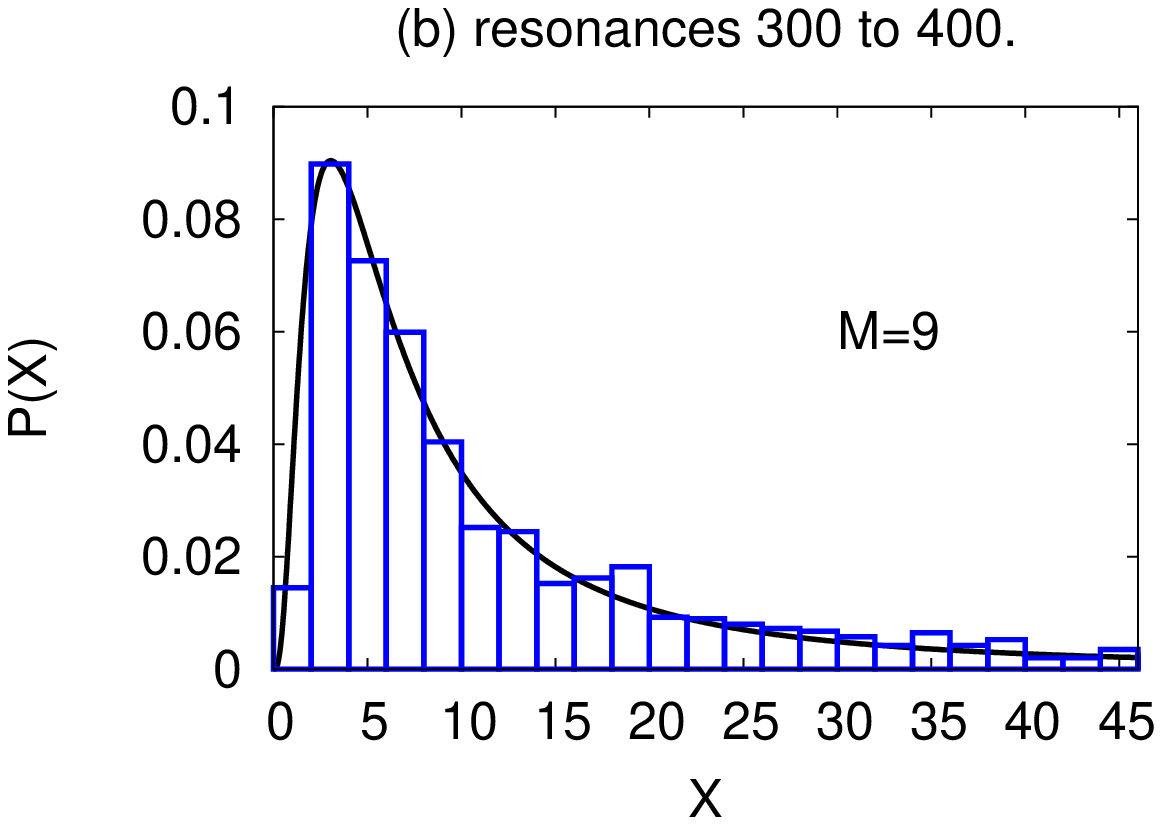}\\
\includegraphics[width=2.5in]{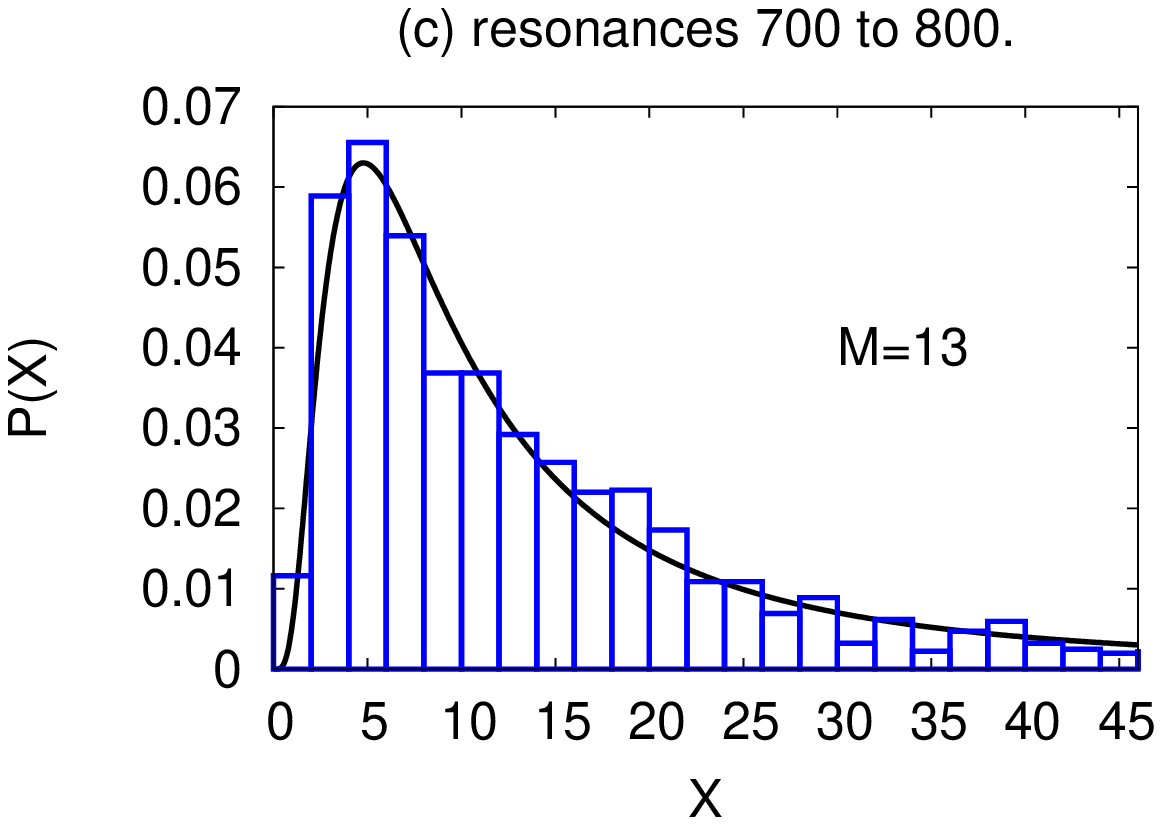}\\
\includegraphics[width=2.5in]{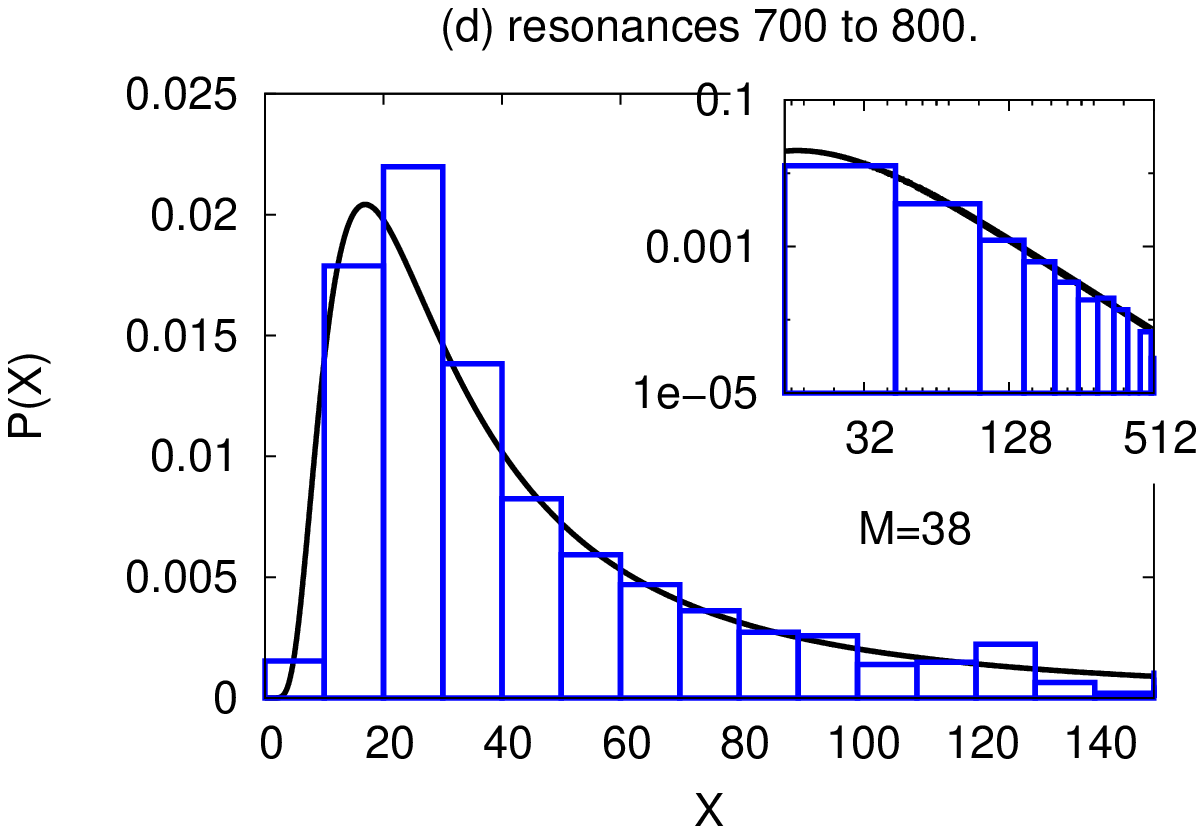}
\end{center} 
\caption{(Color on line). Distributions of the complexness parameter $\Pgoe$ for the same absorbing length, conductivities, frequency ranges  as given in Fig. \ref{pgFig}. The histograms show the numerical distributions. The solid lines correspond to the theoretical prediction (\ref{Mom}) where the number $M$ of channels corresponds to the nearest integer value of relation (\ref{Sab}) computed with the median value of the wavelength in each frequency interval.}
\label{pXFig}
\end{figure}

Thus, having validated the description of losses through the introduction of effective scattering channels, whose number $M$ is solely fixed through the length of the absorbing part of the boundary, the numerical distribution of the complexness parameter $X$ is finally compared to the theoretical prediction (Fig. \ref{pXFig}) for the same absorbing lengths, conductivities, and frequency ranges as those given in Fig. \ref{pgFig}. The excellent agreement, even in the tail of the distribution, confirms that the prediction (\ref{pMX}), obtained within the perturbation theory, contains the essential features to account for the complexness of the resonance states due to spatially continuously distributed losses. 

In conclusion, we investigated the statistics of complex resonance states in an open 2D chaotic microwave cavity by solving the Maxwell equations in a cavity whose boundary is subject to Ohmic losses. We successfully compared the statistics of its resonance states and associated widths with the predictions of an effective non-Hermitian Hamiltonian model. In the limit of weak coupling, we have shown that spatially continuously distributed losses could be mapped to a discrete model involving a finite number $M$ of coupling channels, which constitutes a variable parameter in the cavity. To our knowledge, these results are the first unambiguous confirmation of the adequacy of an effective Random Matrix description to account for the spectral and spatial statistics of an open chaotic wave system with continuously distributed losses. 

\begin{acknowledgments}
We wish to thank   Laurent Labont\'e for his helpful support with Comsol$^\text{TM}$.
\end{acknowledgments}

\clearpage

\clearpage

\end{document}